\RequirePackage{fix-cm}
\documentclass[smallextended]{svjour3}       
\smartqed  
\usepackage{graphicx}
\begin{document}

\title{Relativistic dynamics of oppositely charged two fermions interacting with external uniform magnetic field
}

\titlerunning{Two fermions interacting with external uniform magnetic field}        

\author{Abdullah Guvendi\and Semra Gurtas Dogan}

\authorrunning{A Guvendi et~al.} 

\institute{Medical Imaging Techniques, Simav Vocational School of Health Services, Kutahya Health Sciences University, TR- 43500 Kutahya, Turkey \\
              \email{abdullah.guvendi@ksbu.edu.tr}          
           \and  \email{semragurtas@yahoo.com}  
}

\date{Received: date / Accepted: date}

\maketitle

\begin{abstract}
We investigated the relativistic dynamics of oppositely charged two fermions interacting with an external uniform magnetic field. We chose the interaction of each fermion with the external magnetic field in the symmetric gauge, and obtained a precise solution of the corresponding fully-covariant two-body Dirac equation that derived from Quantum Electrodynamics via Action principle. The dynamic symmetry of the system we deal with allowed us to determine the relativistic Landau levels of such a spinless composite system, without using any group theoretical method. As a result, we determined the eigenfunctions and eigenvalues of the corresponding two-body Dirac Hamiltonian.
\end{abstract}

\section{Introduction}

\qquad Landau quantization is the quantization of cyclotron orbits of the motion of a charged particle moving in a magnetic field \cite{landau1930diamagnetismus}. In the non-relativistic regime, Landau quantization was discussed for many different cases \cite{dunne1992hilbert,paredes20011,ribeiro2006landau}. In the relativistic regime, Landau quantization firstly was discussed by Jackiw \cite{jackiw1984fractional} (see also \cite{balatskii1986chiral}). Afterwards, many experimental and theoretical studies were conducted on this subject \cite{ericsson2001towards,li2007observation,moessner1996exact,furtado1994landau,wang2017valley,fu2020anomalous,Marques,Bueno,netto2008elastic,MAIA2020168229,Amaro Neto,Muniz,Cunta,Figueiredo,bakke2009relativistic,Silva} (also see \cite{Furtado,ribeiro2008landau}). Today, it is though that the magnetic fields exist in all over the spacetime background and they magnetize the universe \cite{enqvist1994ferromagnetic,grasso2001magnetic}. We do not know yet exactly the origins of intra-cluster, galactic and cosmological magnetic fields, but it is predicted that dynamo effects in turbulent fluids can exponentially amplify the seed fields \cite{sigl_1997primordial}. Although it is not easy to fully explain the galactic and cluster magnetic fields with dynamo theory \cite{vainshtein1991turbulent}, it is suggested that these fields that permeate the universe may have arisen as a result of the compression of a primordial field \cite{sigl_1997primordial,gruzinov1996small}. Therefore, the determination of the magnetic field effects at all scales from planets to stars, from galaxies to galaxy clusters and even in the inner galactic environment is still one of the important and active research fields \cite{ryu2008turbulence,schober2020generation}. Also, it has been studied for many years that the structure of the vacuum is modified in the presence of electromagnetic fields \cite{heisenberg1936folgerungen,hattori2013vacuum}. The modification of the vacuum can cause some novel phenomena such as photon decays into electron-positive electron (positron) pair \cite{alkofer2001pair}, the birefringence of a photon \cite{hattori2013vacuum} and splitting of a photon \cite{adler1971photon}. It is natural to expect that these effects can occur, since the vacuum (in QED) is regarded as filled with electrons. Hence, they can react like an ordinary medium in the presence of external fields. We think that magnetic fields exist at almost every point in the universe \cite{kulsrud2008origin} and they may be responsible for many interesting physical effects. The effect illustrated in Nambu-Jona-Lasinio model is that a constant magnetic field is a strong catalyst of a dynamical flavour symmetry breaking \cite{gusynin1994catalysis}. It is though that this point may be important in the phenomena like Hall effect in the condensed matter systems \cite{gusynin1994catalysis}. The presence of magnetic fields can deeply affect the dynamics of relativistic and semi-relativistic systems and hence the determination of the dynamics of systems in magnetic field is a subject of great interest in many areas of physics \cite{miransky2015quantum}. The origin of the magnetic field differs from one system to another, but in some cases we expose the systems to the effect of  external magnetic fields in order to better understand the underlying physics of systems or to determine the effect of the magnetic field on the systems. Of course, the production of magnetic fields can be a natural feature of the systems. In this present manuscript, we will investigate the relativistic dynamics of oppositely charged two fermions interacting with external magnetic field, without detaily discussing the origin of the external magnetic field.

\qquad It is worth to mention that, in general, in the relativistic regime phenomenologically established one-time equations are used to describe the relativistic dynamics of interacting particles. These phenomenologically established equations include free Hamiltonians for each particle plus interparticle interaction potentials \cite{kemmer1937,fermi1949}. It is crucial to mention that there are main difficulties even in the determination of relativistic dynamics of two particles. One of them is that two-time problem appears in such problems in the relativistic framework (see \cite{giachetti2019}). Also, one of the other problems is total angular momentum of the interested composite structures \cite{giachetti2019}. In the literature, the accepted first relativistic two-body equation was introduced by Breit \cite{breit1929}. The history of relativistic two-body equations goes long way back. More details about them can be found in \cite{giachetti2019,van1997}. It is important to note that the equation introduced by Breit does not hold in every situation, due to the retardation effects. Bethe and Salpeter introduced another formalism \cite{salpeter1951} to overcome this problem. The Bethe-Salpeter formalism provided a different approach to one-electron atom problems, but this formalism was not fully-acceptable for bound-states (see \cite{giachetti2019}). Barut and his collaborators have discussed whether it is possible to obtain Poincaré invariant many-body one-time equation \cite{barut1985derivation,barut1986center} (also see \cite{moshinsky1993barut}). They have shown us how is possible to derive a complete fully-covariant two-body Dirac equation \cite{barut1985derivation}. This equation has been derived from Quantum Electrodynamics with the help of Action principle \cite{barut1985derivation}. This equation is very similar to the former two-body equation introduced by Kemmer, Fermi and Yang \cite{kemmer1937,fermi1949} and moreover it is in fully-covariant form (more details can be found in Refs. \cite{barut1985derivation,barut1986center,moshinsky1993barut,barut1985,barut1987a,barut1987b,guvendi2019exact}). In $3+1$ dimensions, the fully-covariant two-body equation gives a $16\times16$ dimensional matrix equation including the most general electric and magnetic potentials. But, in $4$-dimensions, the solution of this equation requires to be applied group theoretical methods \cite{barut1985} and the well-known energy spectrum for Hydrogen-like atoms can be obtained only via a perturbative solution of this equation \cite{barut1985,barut1987a,barut1987b}. Nevertheless, it is showed that the fully-covariant two-body Dirac equation can be exactly solvable for composite systems that have dynamical symmetries \cite{guvendi2019exact}. On the other hand, one can see that it may not be possible to obtain a complete analytical solution for one-electron atom systems under the influence of external magnetic field with the corresponding one-body Dirac equation \cite{de2014bound} even in the absence of the third spatial coordinate \cite{chiang2002planar,khalilov1998dirac}. In the literature, one can see that there are a few study based on relativistic dynamics of non-interacting (mutually) two particles in the presence of external magnetic field. Previously, the separation of center of mass motion coordinates of a two-body system (neutral) in homogeneous magnetic fields was introduced \cite{avron1978separation} and then the theoretical foundations of two-body systems were studied in the presence of both homogeneous \cite{herold1981two} and inhomogeneous \cite{bock2005neutral} magnetic fields (also see \cite{gorbar2015supercritical,gorbar2015supercriticality}).

\qquad In this present paper we hoped to fill this gap in the literature and we investigated the relativistic dynamics of oppositely charged two fermions interacting with an external uniform magnetic field, without considering any mutually charge-charge interaction. At the beginning, we wrote the fully-covariant two body Dirac equation in $3$-dimensional Minkowski spacetime \cite{guvendi2019exact} and we chose the interaction of each fermion with the external magnetic field in the symmetric gauge, since this problem has $2+1$-dimensional dynamical symmetry \cite{semra2019}. The dynamic symmetry of the system allowed us to determine the eigenfunctions and eigenvalues (in closed-form) of the corresponding two-body Dirac Hamiltonian for such a spinless composite system, without using any group theoretical method.

\section{Fully-covariant two-body Dirac equation in 2+1 dimensions}
\label{sec:1}

\qquad In a general three dimensional spacetime background, the relativistic dynamics of charged two fermions interacting with external uniform magnetic field can be investigated via the following fully-covariant two-body Dirac equation,
\begin{eqnarray}
&\left\lbrace\left[\gamma^{\eta^{\left(1\right)}}\pi^{\left(1\right)}_{\eta}+ib_{1}\textbf{I}_{2}\right]\otimes\gamma^{0^{\left(2\right)}}+\gamma^{0^{\left(1\right)}}\otimes\left[\gamma^{\eta^{\left(2\right)}}\pi^{\left(2\right)}_{\eta}+ib_{2}\textbf{I}_{2}\right]  \right\rbrace \Psi\left(\textbf{x}_{1},\textbf{x}_{2}\right)=0,\nonumber\\
&\pi^{\left(1\right)}_{\eta}=\left(\partial^{\left(1\right)}_{\eta}+i\frac{e_{1}A^{\left(1\right)}_{\eta}}{\hbar c}-\Gamma_{\eta}^{\left(1\right)}\right),\quad \pi^{\left(2\right)}_{\eta}=\left(\partial^{\left(2\right)}_{\eta}+i\frac{e_{2}A^{\left(2\right)}_{\eta}}{\hbar c}-\Gamma_{\eta}^{\left(2\right)}\right),\nonumber\\
&b_{1}=\frac{m_{1}c}{\hbar},\quad b_{2}=\frac{m_{2}c}{\hbar},\quad \left(\eta=0,1,2.\right),\label{Eq1}
\end{eqnarray}
in which the superscripts $(1) $ and $(2)$ refer to the first fermion with the mass $m_{1}$ and second fermion with mass $m_{2}$, respectively. In Eq. (\ref{Eq1}), $\textbf{I}_{2}$ is $2\times2$ dimensional unit matrix, $\Psi \left( \mathbf{x}_{1},\mathbf{x}_{2}\right)$ is the composite field that is constructed by a direct production $\left(\otimes\right)$ of arbitrary massive two Dirac fields as follows,
\begin{eqnarray}
\Psi \left( \mathbf{x}_{1},\mathbf{x}_{2}\right) =\Upsilon \left( \mathbf{x}_{1}\right)\otimes\chi \left(\mathbf{x}_{2}\right),\label{Eq2}
\end{eqnarray}
$e_{1}$ and $e_{2}$ are charges of these particles, $\hbar$ is usual Planck constant, the letter $c$ represents to the light speed, $A_{\eta }$ and $\Gamma_{\eta}$ correspond to the vector potentials and spinor connections, respectively. It has been showed that this two-body Dirac equation is Poincaré invariant \cite{barut1985derivation,barut1986center,moshinsky1993barut} and one-time equation including spin algebra spanned by direct (Kronocker) productions of the Dirac matrices. Even though Eq. (\ref{Eq1}) does not seem to be manifestly covariant at first look, the $\gamma^{0}$ means $\gamma^{\eta}\lambda_{\eta}$ in everywhere in this equation (note that $\lambda_{\eta}$ is a timelike vector $\lambda_{\eta}=(100)$ and more details about its structure can be found in Refs. \cite{barut1985derivation,barut1986center,moshinsky1993barut,barut1985,barut1987a,barut1987b,guvendi2019exact}). As we mentioned in above, the problem we deal with can be studied in $2+1$ dimensional flat Minkowski spacetime background that can be represented via the following line element,
\begin{eqnarray}
ds^{2}=c^{2}dt^{2}-dx^{2}-dy^{2}.\label{Eq3}
\end{eqnarray}
It is clear that the relativistic dynamics of the system we deal with does not changed by spinor connections in Eq. (\ref{Eq1}), since they vanish \cite{sucu2007}. It is thought that there exist magnetic field at almost in every point in the universe and the vacuum (in Quantum Electrodynamics) is filled by fermions (electron/positron). Hence, without considering any charge-charge interaction, we can think that the relativistic dynamics of a fermion-antifermion system in the presence of external magnetic field thought to be important to better understand the universe. Therefore, in the presence of the external uniform magnetic field, the vector potentials in Eq. (\ref{Eq1}) can be chosen in the symmetric gauge as follows (without any charge-charge interaction term, see also \cite{guvendi2019exact}),
\begin{eqnarray}
&A^{\left(1\right)}_{0}=0,\quad A^{\left(1\right)}_{1}=-\frac{B_{0}y_{1}}{2}, \quad A^{\left(1\right)}_{2}=\frac{B_{0}x_{1}}{2},\nonumber\\
&A^{\left(2\right)}_{0}=0,\quad A^{\left(2\right)}_{1}=-\frac{B_{0}y_{2}}{2}, \quad A^{\left(2\right)}_{2}=\frac{B_{0}x_{2}}{2},\label{Eq4}
\end{eqnarray}
in which $B_{0}$ relates with the strength of external uniform magnetic field and $x_{p},y_{p} \left(p=1,2.\right)$ pairs correspond to the spatial coordinates of the particles in the spacetime background represented by Eq. (\ref{Eq3}).
For the line element given in Eq. (\ref{Eq3}), the Dirac matrices can be chosen as in the following \cite{guvendi2019exact},
\begin{eqnarray}
&\gamma^{0^{\left(1,2\right)}}=\sigma^{z},\quad \gamma^{1^{\left(1,2\right)}}=i\sigma^{x},\quad \gamma^{2^{\left(1,2\right)}}=i\sigma^{y},\nonumber\\
&\sigma^{x}=\left(
\begin{array}{cc}
0 & 1 \\
1  &0\\
\end{array}
\right),\quad \sigma^{y}=\left(
\begin{array}{cc}
0 & -i \\
i  &0\\
\end{array}
\right),\nonumber\\
&\quad \sigma^{z}=\left(
\begin{array}{cc}
1 & 0 \\
0  &-1\\
\end{array}
\right),\label{Eq5}
\end{eqnarray}
where $\sigma^{x}$, $\sigma^{y}$ and $\sigma^{z}$ are the Pauli spin matrices.

\section{Radial equations}
\label{sec:2}

\qquad As is usual with two-body problems, the center of mass motion coordinates and relative motion coordinates can be separated (covariantly) via help of the following expressions \cite{guvendi2019exact},
\begin{eqnarray}
&R_{\eta}=\frac{1}{M}\left(m_{1}x_{\eta}^{\left(1\right)}+m_{2}x_{\eta}^{\left(2\right)}\right),\quad r_{\eta}=x_{\eta}^{\left(1\right)}-x_{\eta}^{\left(2\right)},\nonumber\\
&x_{\eta}^{\left(1\right)}=\frac{m_{2}}{M}r_{\eta}+R_{\eta},\quad x_{\eta}^{\left(2\right)}=-\frac{m_{1}}{M}r_{\eta}+R_{\eta},\nonumber\\
&\partial_{x_{\eta}}^{\left(1\right)}=\partial_{r_{\eta}}+\frac{m_{1}}{M}\partial_{R_{\eta}},\quad \partial_{x_{\eta}}^{\left(2\right)}=-\partial_{r_{\eta}}+\frac{m_{2}}{M}\partial_{R_{\eta}},\nonumber\\
&\partial_{x_{\eta}}^{\left(1\right)}+\partial_{x_{\eta}}^{\left(2\right)}=\partial_{R_{\eta}}.\label{Eq6}
\end{eqnarray}
It is important to underline that the total energy of the system is determined according to the proper time $(R_{0})$ of the system, since there is no relative time difference between the first and second fermions. Now, we can assume that the center of mass is located at the origin of the spacetime background ($R_{1}=R_{2}=0$). At that rate, by substituting Eq. (\ref{Eq4}), Eq. (\ref{Eq5}) and Eq. (\ref{Eq6}) into Eq. (\ref{Eq1}) one can acquire the following matrix equation,
\begin{eqnarray}
&\left(\gamma^{0}\otimes\gamma^{0}\right)\partial_{R_{0}}\Psi +i \frac{m_{1}c}{\hbar}\left(\textbf{I}_{2}\otimes\gamma^{0}\right)\Psi+i \frac{m_{2}c}{\hbar}\left(\gamma^{0}\otimes\textbf{I}_{2}\right)\Psi\nonumber\\
&+\left(\gamma^{1}\otimes\gamma^{0}\right)\left(\partial_{x}+\frac{m_{1}}{M}\partial_{X}-iB_{1}\frac{m_{2}}{M}y\right)\Psi+\left(\gamma^{2}\otimes\gamma^{0}\right)\left(\partial_{y}+\frac{m_{1}}{M}\partial_{Y}+iB_{1}\frac{m_{2}}{M}x\right)\Psi\nonumber\\
&\resizebox{.99\hsize}{!}{$+\left(\gamma^{0}\otimes\gamma^{1}\right)\left(-\partial_{x}+\frac{m_{2}}{M}\partial_{X}+iB_{2}\frac{m_{1}}{M}y\right)\Psi+\left(\gamma^{0}\otimes\gamma^{2}\right)\left(-\partial_{y}+\frac{m_{2}}{M}\partial_{Y}-iB_{2}\frac{m_{1}}{M}x\right)\Psi=0$},\nonumber\\
&B_{1}=\frac{e_{1}B_{0}}{2\hbar c},\quad B_{2}=\frac{e_{2}B_{0}}{2\hbar c}, \label{Eq7}
\end{eqnarray}
in which $x,y$ and $X,Y$ pairs are the spatial coordinates of the relative motion and center of mass motion, respectively. Provided that center of mass momentum is a constant of motion and the interaction is time-independent, we can define the composite field $\Psi$ as in the following,
\begin{eqnarray}
&\Psi\left(t,\textbf{r},\textbf{R}\right)= e^{-i\omega t}e^{i\textbf{K}.\textbf{R}}\Omega\left(\textbf{r}\right),\quad \Omega\left(\textbf{r}\right)=\left(
\begin{array}{c}
\psi_{1}\left(\textbf{r}\right)  \\
\psi_{2}\left(\textbf{r}\right)  \\
\psi_{3}\left(\textbf{r}\right)  \\
\psi_{4}\left(\textbf{r}\right)
\end{array}
\right),\label{Eq8}
\end{eqnarray}
in which the $\omega$ is total frequency determined according to the proper time of the system and the $\textbf{K}$ relates with the spatial momentum of the center of mass motion ($\hbar\textbf{K}$). For the ansatz in Eq. (\ref{Eq8}) defined for a moving $(\textbf{K}\neq0)$ system formed by oppositely charged $(e_{1}=e,e_{2}=-e)$ and arbitrary massive two fermions, by multiplying the Eq. (\ref{Eq7}) with $\gamma^{0}\otimes\gamma^{0}$ from left \footnote{Here, $\left(\gamma^{0}\otimes \gamma^{0}\right)^{2}$ gives $4\times4$ dimensional unit matrix.}, one can obtain the following matrix equation,
\begin{eqnarray}
&\resizebox{.99\hsize}{!}{$\left(
\begin{array}{cccc}
\varepsilon-M & \partial _{-} &
-\partial _{-} & 0 \\
-\partial _{+}& \varepsilon-\Delta m & 0
& -\partial _{-} \\
\partial _{+} & 0 & \varepsilon+\Delta m &
\partial _{-} \\
0 & \partial _{+} & -
\partial _{+} & \varepsilon+M
\end{array}
\right) \Omega\left(\textbf{r}\right)-\left(
\begin{array}{cccc}
0 & \mu_{1}B r_{-} &
\mu_{1}iK_{-}& 0 \\
 \mu_{1}B r_{+} & 0 & 0
&\mu_{1}iK_{-} \\
-\mu_{1}iK_{+}  & 0 & 0 &
\mu_{1}B r_{-} \\
0 & -\mu_{1}iK_{+} &\mu_{1}B r_{+}  & 0
\end{array}
\right)\Omega\left(\textbf{r}\right)$}\nonumber\\
&-\left(
\begin{array}{cccc}
0 &\mu_{2}iK_{-} &
\mu_{2}B r_{-} & 0 \\
-\mu_{2}iK_{+} & 0 & 0
& \mu_{2}B r_{-} \\
\mu_{2}B r_{+}  & 0 & 0 &
\mu_{2}iK_{-}  \\
0 &\mu_{2}B r_{+}&\mu_{2}-iK_{+}  & 0
\end{array}
\right)\Omega\left(\textbf{r}\right)=0,\nonumber\\
&M=\frac{\left(m_{1}+m_{2}\right)c}{\hbar},\quad \Delta m=\frac{\left(m_{1}-m_{2}\right)c}{\hbar},\quad B=\frac{eB_{0}}{2\hbar c}\nonumber\\
&\varepsilon=\frac{\omega}{c},\quad \partial _{\pm}=\partial _{x}\pm i\partial _{y},\quad r_{\pm}=x\pm iy,\nonumber\\
&K_{\pm}=K_{x}\pm iK_{y},\quad \mu_{q}=m_{q}/(m_{1}+m_{2}),\quad \left(q=1,2.\right). \label{Eq9}
\end{eqnarray}
\qquad We can transform the spacetime background into the polar spacetime background in order to exploit the angular symmetry. We can do this by using the following expressions,
\begin{eqnarray*}
\partial _{\pm} = e^{\mp i\phi }\left(\mp
 \frac{i}{r}\partial _{\phi }+\partial _{r}\right),\quad x\pm iy=re^{\pm i\phi },\quad K_{\pm}=Ke^{\pm i\phi },
\end{eqnarray*}
in which $\partial _{-}$ and $\partial _{+}$ represent to the well-known spin lowering and spin raising operators, respectively. Now, we can write the matrix equation in Eq. (\ref{Eq9}) in terms of polar coordinates $(r,\phi)$ only for the transformed spinor components that are defined as follows,
\begin{eqnarray*} \left(
\begin{array}{c}
\psi _{1}\left( \textbf{r}\right)  \\
\psi _{2}\left( \textbf{r}\right)  \\
\psi _{3}\left( \textbf{r}\right)  \\
\psi _{4}\left( \textbf{r}\right)
\end{array}
\right)
\Longrightarrow\left(
\begin{array}{c}
\cr \psi _{1}\left(r\right)e^{i\left(s-1\right)\phi}  \\
\cr \psi _{2}\left(r\right)e^{is\phi}  \\
\cr \psi _{3}\left(r\right)e^{is\phi}  \\
\cr \psi _{4}\left(r\right)e^{i\left(s+1\right)\phi}
\end{array}
\right),
\end{eqnarray*}
as,
\begin{eqnarray}
&\varepsilon\varphi_{1}\left(r\right)-M\varphi_{2}\left(r\right)+\left(2\partial_{r}+i\frac{\Delta m}{M}K \right)\varphi_{3}\left(r\right)-B r\varphi_{4}\left(r\right)=0,\nonumber\\
\nonumber\\
&\varepsilon\varphi_{2}\left(r\right)-M\varphi_{1}\left(r\right)+\left(\frac{2s}{r}-B \frac{\Delta m}{M}r \right)\varphi_{3}\left(r\right)-iK\varphi_{4}\left(r\right)=0,\nonumber\\
\nonumber\\
&\varepsilon\varphi_{3}\left(r\right)-\Delta m\varphi_{4}\left(r\right)+\left(\frac{2s}{r}-B\frac{\Delta m}{M}r \right)\varphi_{2}\left(r\right)-\left(\frac{2}{r}+2\partial_{r}+i\frac{\Delta m}{M}K\right)\varphi_{1}\left(r\right)=0,\nonumber\\
\nonumber\\
&\varepsilon\varphi_{4}\left(r\right)-\Delta m\varphi_{3}\left(r\right)-B r\varphi_{1}\left(r\right)+iK\varphi_{2}\left(r\right)=0,\label{Eq10}
\end{eqnarray}
in which,
\begin{eqnarray*}
&\varphi_{1}\left(r\right)=\psi_{1}\left(r\right)+\psi_{4}\left(r\right),\quad \varphi_{2}\left(r\right)=\psi_{1}\left(r\right)-\psi_{4}\left(r\right),\nonumber\\
&\varphi_{3}\left(r\right)=\psi_{2}\left(r\right)-\psi_{3}\left(r\right),\quad \varphi_{4}\left(r\right)=\psi_{2}\left(r\right)+\psi_{3}\left(r\right),
\end{eqnarray*}
and the letter $s$ stands for the total spin of system formed by two fermions. There are two main difficulties in a complete analytical solution of this system of coupled equations. The first difficulty is the mass difference between the particles even for the static case ($K=0$). The second difficulty is the total angular momentum of such a system as we mentioned before \cite{giachetti2019}. For such a spinless $(s=0)$ system formed by equal massive ($m_{1}=m_{2}=m$) oppositely charged two fermions, one can obtain the following system of coupled equations by defining a dimensionless independent variable that reads $z=\frac{B}{2}r^{2}$,
\begin{eqnarray}
&\varepsilon\varphi_{1}\left(z\right)-M\varphi_{2}\left(z\right)+2\sqrt{\frac{2z}{B}}B\partial_{z}\varphi_{3}\left(z\right)-\sqrt{\frac{2z}{B}}B\varphi_{4}\left(z\right)=0,\nonumber\\
\nonumber\\
&\varepsilon\varphi_{2}\left(z\right)-M\varphi_{1}\left(z\right)=0,\nonumber\\
\nonumber\\
&\varepsilon\varphi_{3}\left(z\right)-\sqrt{\frac{2B}{z}}\varphi_{1}\left(z\right)-2\sqrt{\frac{2z}{B}}B\partial_{z}\varphi_{1}\left(z\right)=0,\nonumber\\
\nonumber\\
&\varepsilon\phi_{4}\left(z\right)-\sqrt{\frac{2z}{B}}B\varphi_{1}\left(z\right)=0.\label{Eq11}
\end{eqnarray}
in the rest frame  $(K=0)$. Of course, the $K=0$ case is relatively simple case, but any pairing effect becomes important in this case.

\section{Energy spectrum}
\label{sec:3}

\begin{figure}
 \includegraphics{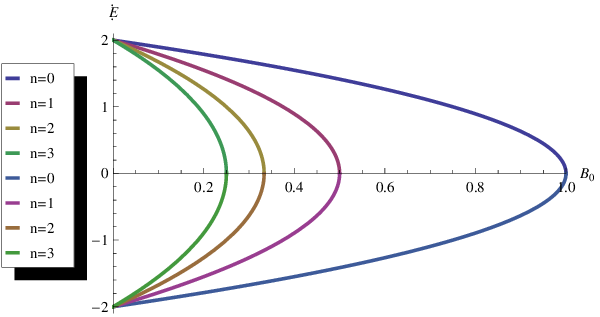}
\caption{The dependence of total energy on the strength of external uniform magnetic field. Here, $|e|=m=c=\hbar=1$.}
\label{fig:1}
\end{figure}

\qquad One can solve the system of coupled equations in Eq. (\ref{Eq11}) in favour of $\varphi_{1}\left(z\right)$ and arrive at the following $2^{nd}$ order wave equation,
\begin{eqnarray*}
\partial^{2}_{z}\varphi_{1}\left(z\right)+\frac{1}{z}\partial_{z}\varphi_{1}\left(z\right)-\frac{1}{4}\left(\frac{z^{2}+1}{z^{2}}-\frac{\varepsilon^{2}-M^{2}}{2B z}\right)\varphi_{1}\left(z\right)=0,\label{Eq12}
\end{eqnarray*}
This equation can be reduced into the well-known shape of the Whittaker differential equation via defining the ansatz that reads $\varphi_{1}\left(z\right)=\frac{1}{\sqrt{z}}\chi\left(z\right) $,
\begin{eqnarray}
&\partial^{2}_{z}\chi\left(z\right)+\left(\frac{\mu}{z}+\frac{\frac{1}{4}-\nu^{2}}{z^{2}}-\frac{1}{4}\right)\chi\left(z\right)=0,\quad \mu =\frac{\varepsilon^{2}-M^{2}}{8 B},\quad \nu=\frac{1}{2},\label{Eq13}
\end{eqnarray}
and the solution function of this wave equation is given in follows \cite{book:874771,dernek2018relativistic},
\begin{eqnarray*}
&\chi\left(z\right)=QW_{\mu,\nu}(z )
\end{eqnarray*}
in which the $Q$ is the normalization constant. The condition of the solution function to be polynomial is given as follows \cite{book:874771,dernek2018relativistic},
\begin{eqnarray}
\frac{1}{2}+\nu-\mu=-n,\label{Eq14}
\end{eqnarray}
in which the $n$ is principal quantum number (non-negative integer). The expression in Eq. (\ref{Eq14}) leads to the quantization condition for the formation of such a spinless system. With the help of Eq. (\ref{Eq13}) and Eq. (\ref{Eq14}) one can acquire the following non-perturbative spectrum in energy domain by assuming $e<0$,
\begin{eqnarray}
E=\pm 2mc^{2}\sqrt{1-\frac{\omega_{c}\hbar}{mc^{2}}\left(n+1\right)},\quad \omega_{c}=\frac{|e|B_{0}}{m c},\label{Eq15}
\end{eqnarray}
in which $\omega_{c}$ is the well-known cyclotron frequency. Eq. (\ref{Eq15}) clearly gives the relativistic Landau levels of such a system (spinless) formed by oppositely charged two fermions (non-interacting). Dependence of total energy ($E$) on the strength of the external homogeneous magnetic field can be seen in Fig. \ref{fig:1}. Also, one can obtain all of the spinor components as follows,

\begin{eqnarray}
\left(
\begin{array}{c}
\varphi_{1}\left(z\right) \\
\varphi _{2}\left(z\right) \\
\varphi _{3}\left(z\right) \\
\varphi _{4}\left(z\right)
\end{array}
\right) =Q\left(
\begin{array}{c}  \frac{W_{\mu ,\nu}(z )}{\sqrt{z}}
\\ \frac{M}{\varepsilon \sqrt{z}}W_{\mu ,\nu}(z ) \\
\frac{\left(z-2\mu\right)\sqrt{2B}}{\varepsilon z}W_{\mu ,\nu}(z )-\frac{2\sqrt{2B}}{\varepsilon z}W_{\mu+1 ,\nu}(z ) \\
\frac{\sqrt{2B}}{\varepsilon}W_{\mu ,\nu}(z )\end{array}\right),\label{Eq16}
\end{eqnarray}
which, of course, satisfy the system of coupled equations given in Eq. (\ref{Eq11}).

\section{Results and Discussion}
\label{sec:4}

\qquad In this study, we investigated the relativistic dynamics of oppositely charged two fermions interacting with an external homogeneous magnetic field. To obtain a non perturbative energy spectrum of such a system, we solved the corresponding fully-covariant two-body Dirac equation in $2+1$ dimensional flat Minkowski spacetime background, since this problem has $2+1$-dimensional dynamical symmetry. For a spinless and static system formed by oppositely charged two fermions (non-interacting), the dynamic symmetry of the problem we deal with allowed us to obtain the eigenfunctions and eigenvalues (in closed-form) of the corresponding fully-covariant two-body Dirac Hamiltonian, without using any group theoretical method. As it expected, the obtained energy spectrum (Eq. (\ref{Eq15})) shows that the total energy ($E$) of the system closes to total rest mass energy ($2mc^{2}$) of the system when the external magnetic field is very weak $(\omega_{c}\hbar\ll mc^{2})$. It is clear that in Eq. (\ref{Eq15}), the term associated with the external magnetic field does not vanish even for the ground state ($n=0$) of such a system. Also, the total energy value of the system closes to the zero when $\omega_{c}\hbar\approx mc^{2}$ and $n=0$. The non-perturbative energy spectrum in Eq. (\ref{Eq15}) also indicates that this system can decay when $\omega_{c}\hbar > mc^{2}$ in any physically possible quantum state.

\begin{acknowledgements}
The authors thank Dr. Yusuf SUCU for suggestions and useful discussions and anonymous referee for valuable comments and style suggestions.
\end{acknowledgements}

\end{document}